# Quantum Cryptography: where do we stand?

Nicolas Gisin
Group of Applied Physics
University of Geneva - Switzerland

In the last few years the world in which Quantum Cryptography evolves has deeply changed. On the one side the revelations of Snowden, though he said nothing really new, made the world more aware of the importance to protect sensitive data from all kinds of adversaries, including sometimes "friends". On the other side, breakthroughs in quantum computation, in particular in superconducting qubits and surface-codes, made it possible that in 15 to 25 years there might be a quantum machine able to break today's codes. This implies that in order to protect today's data over a few decades, one has to act now and use some quantum-safe cryptography.

Such a perspective is taken very seriously by the cryptography community who is looking for a quantum-safe alternative to today's systems.

Quantum-safe cryptography covers all cryptography systems that resist to quantum attacks. As in today's cryptography, this covers both complexity-based protocols and provably secure systems. The first ones consists in merely replacing the problem on which RSA rests (i.e. factoring) by another problem claimed to be intractable both for classical and for quantum computers, as factoring was claimed to be intractable. This approach has the great advantage of being flexible, cost-effective and relatively similar to today's approach, hence security experts don't need to change much. But it has the great drawback that one is again betting on the unknown to secure our information-based society. Provably secure cryptography requires that one exploit physics, more precisely quantum physics, a relatively expensive and different approach, but a real paradigm shift.

Complexity-based protocols, like e.g. lattice based encryption, will certainly find large markets in everyday applications. Yet, even for this low end of the market, quantum technologies have an important role to play. Indeed, all cryptography protocols require fresh random numbers and Snowden clearly established that one can't count on pseudo-random number generators. Quantum offers guaranteed randomness. This can be made very handy, compact and cost effective. It would be absurd not to use Quantum Random Number Generators (QRNG) in all future cryptography applications.

Provably secure cryptography combines a physical layer for Quantum Key Distribution (QKD), QRNG and a large part of classical algorithms to provide ultra-long time confidentiality. Here one should always emphasize that a glitch in complexity based cryptography break not only future communication, but also all past communication, while QKD-based systems can only be attacked in real time. Hence, a glitch on a QKD-based system does not at all affect past communication and its effect on future communication depends on the detail of the hypothetical glitch. For example, if authentication is realized by a classical one-way function, then even if it the one-way function gets broken after a minute, this has no impact on the security since after a minute the QKD-based system has already gone on with a fresh round of bases choices.

Quantum networks using Quantum Key Distribution and Trusted Nodes are being deployed in several countries (USA, China, Korea, UK, etc.). They provide a solid physical backbone on which to develop quantum-safe cryptography, especially for high security and long-term secrecy. Trusted node is a compromise. Today's QKD is limited to a few hundreds of km [1], despite steady progress in reducing losses of optical fibers and in improving QKD protocols and engines. Quantum repeaters based on quantum teleportation and memories still need to be much improved before they will be useful for country-scale quantum networks (how long depends on the physicists creativity and the engineers budget).

As QKD-based systems get more and more deployed, testing those systems becomes more and more important. Indeed, even if the principle of QKD is provably secure, implementation may have weaknesses. Let us emphasize again that the discovery of a weakness would not impact the security of past communication. I like to also emphasize that most – possibly all – weaknesses discovered so far, in particular about single-photon detectors, are essentially science-fiction attacks that require that the adversary has access during several hours to the inside of the QKD engine, so that he can fine tune some delicate parameters of his attack. Nevertheless, it is important to know about such attacks. The attacks exploiting the detectors are somewhat similar to those exploiting the photon-number Poisson distribution of the source: both can easily be counters by similar simple counter measures, decoy state for the latter and decoy detectors for the former. For example a well-implemented single-photon detector with randomly varying detection efficiencies (at least 3 levels are required) allows one to mitigate attacks on the detectors [2]. Still, one should recognize that this is not a trivial task [3].

Research in QKD is still needed in three quite different directions. First, there is a need to improve the QKD protocols (including security proofs against individual and collective attacks) and the QKD engines, especially with a view on higher bit rates. A vision of a QKD engine producing 1 Gb/s of provably secret bits is on the horizon. Next, there is a beautiful research program on so-called Device Independent Quantum Information Processing (DIQIP) [4]. This connects to the fascinating physics of quantum nonlocality [5,6], but one should also focus on semi-DIQIP, i.e. on protocols where well-identified parts of the system are trusted. The third main research topic implies very nice experimental physics on light-matter interactions at the single-photon level for the development of quantum memories and repeaters. Each individual requirement for a quantum repeater (e.g. memory time, fidelity, efficiency, etc.) has been demonstrated, but in different and so far incompatible systems.

Hence, the grand challenge here is, first, to find one system able to achieve all requirements and, next, to coordinate a large engineering program to assemble it into a functional quantum repeater for future continental scale quantum networks.